\documentclass[twocolumn, aps]{revtex4}
\usepackage{natbib}

\ifx\pdfoutput\undefined
\usepackage{graphicx}
\else
\usepackage[pdftex]{graphicx}
\usepackage{epstopdf}
\fi

\begin{document}

\title{Collective evolution and the genetic code}
\author{Kalin Vetsigian}

\affiliation{Department of Physics,
University of Illinois at Urbana-Champaign, 1110 West Green St., Urbana, IL 61801}

\author{Carl Woese}
\affiliation{Department of Microbiology and Institute for Genomic
Biology, University of Illinois at Urbana-Champaign, 601 South Goodwin Avenue
Urbana, IL 61801}

\author{Nigel Goldenfeld}
\affiliation{Department of Physics and Institute for Genomic Biology,
University of Illinois at Urbana-Champaign, 1110 West Green St., Urbana, IL 61801}

\date{\today}

\begin{abstract}

A dynamical theory for the evolution of the genetic code is presented,
which accounts for its universality and optimality.  The central
concept is that a variety of collective, but non-Darwinian, mechanisms
likely to be present in early communal life generically lead to
refinement and selection of innovation-sharing protocols, such as the
genetic code.  Our proposal is illustrated using a simplified computer
model, and placed within the context of a sequence of transitions that
early life may have made, prior to the emergence of vertical descent.

\end{abstract}

\maketitle

The genetic code could well be optimized to a greater extent than
anything else in biology, yet is generally regarded as the biological
element least capable of evolving.

There would seem to be four reasons for this paradoxical situation, all
of which reflect the reductionist molecular perspective that so shaped
biological thought throughout the 20th century.  First, the basic
explanation of gene expression appears to lie in its {\it
evolution\/}, and not primarily in the specific structural or
stereochemical considerations that are sufficient to account for gene
replication.  Second, the problem's motto, \lq\lq genetic code", is a
misnomer that makes the codon table the defining issue of gene
expression.  A satisfactory level of understanding of the gene should
provide a unifying account of replication and expression as two sides
of the same coin.  The genetic code is merely the linkage between these
two facets. Thus, and thirdly, the assumption that the code and the
decoding mechanism are separate problems, individually solvable, is a
reductionist fallacy that serves to deny the fundamental biological
nature of the problem.  Finally, the evolutionary dynamic that gave
rise to translation is undoubtedly non-Darwinian, to most an
unthinkable notion that we now need to entertain seriously. These four
considerations structure the approach we take in this paper.

To this point in time, biologists have seen the universality of the
code as either a manifestation of the Doctrine of Common Descent or
simply as a \lq\lq frozen accident".  Viewing universality as following
from common descent renders unthinkable the notion explored here that a
universal code may be a necessary pre-condition for common
ancestry---indeed even for life as we know it.  We will argue in this
paper---a maturation of the earlier concept of the
progenote\cite{WOES77}---that the very fact of the code's evolvability,
together with the details of its internal structure, provide strong
clues to the nature of early life, and in particular its essential
communal character\cite{WOES82}.

Beyond the code's universality we have very few clues to guide us in
trying to understand its evolution and that of the underlying decoding
mechanism. The principal ones again are properties of the code itself;
specifically, the obvious structure of the codon table.  The table
possesses (at least) two types of order: synonym order and relatedness
order.  The first is the relatedness of codons assigned to the same
amino acid; the second the relatedness of codons assigned to related
amino acids. Relatedness among the amino acids is context dependent,
and in the context of the codon table could {\it a priori\/} reflect
almost anything about the amino acids: their various properties, either
individually or in combination; the several macromolecular contexts in
which they are found, such as protein structure, the translation
mechanism, the evolution of translation; or the pre-translational
context of the so-called RNA-world. Although we don't know what defines
amino acid \lq\lq similarity" in the case of the code, we do know one
particular amino acid measure that seems to express it quite remarkably
in the coding context. That measure is amino acid polar
requirement\cite{Woese1965, Woese1966, WOES-book}. While the
relatedness order of the code is marginally evident from simple
inspection of the codon table\cite{Woese1965, Sonneborn, Woese1966,
Alff-Steinberger1969, WONG80}, it is pronounced when the amino acids
are represented by their respective polar requirements\cite{Woese1966}.

A major advance was provided by computer simulation
studies\cite{Haig1991, One_in_a_million, GOLD93,
Freeland2000,Knight-Nat_Rev_Genet,Knight_PhD} of the relatedness
ordering of the amino acids over the codon table, which showed that the
code is indeed relationally ordered, and moreover is optimized to near
the maximum extent possible. Compared to randomly-generated codes, the
canonical code is \lq\lq one in a million" when the relatedness measure
is the polar requirement.  No other amino acid measure is known to
possess this characteristic\cite{Knight_PhD} (in our opinion, the
significance of this observation has not been adequately recognized or
pursued). These precisely-defined relatedness constraints in the codon
table were unexpected, and still cry out for explanation.

As far as interpretation goes, the optimal aspect of the genetic code
is surely a reflection of the last aspect of the coding problem that
needs to be brought into consideration: namely, the precision or
biological specificity with which translation functions. Precision,
along with every aspect of the genetic code, needs to be understood as
part of an evolutionary process.  We would contend that at early stages
in cellular evolution ambiguous translation was tolerated (there being
no alternative), and was an important and essential part of the
evolutionary dynamic (see below).  What we imply by ambiguity here is
inherent in the concept of group codon assignments, where a group of
related codons is assigned as a whole to a corresponding group of
related amino acids\cite{Woese1965}. From this flows the concept of a
\lq\lq statistical protein", wherein a given gene can be translated not
into a unique protein, but instead into a family of related protein
sequences. Note that we do not say that these are an approximation to a
perfect translation of the gene, thereby implying that these sequences
are in some sense erroneous.  Early life did not require a refined
level of tolerance, and so there was no need for a perfect translation.
Ambiguity is therefore not the same thing as \lq\lq error".

The phylogenetic expression of ambiguity is reticulate evolution.  In
reticulate evolution, there is no unique notion of genealogical
descent: genetic content can be distributed collectively. Accordingly,
as we now turn the emphasis away from the documentation of the static
features of the genetic code, and towards their evolutionary origins,
we must necessarily invoke an evolutionary dynamic distinct from that
identified originally by Darwin. This dynamic can be seen as a kind of
biological game in which both the players and the rules of play are
unfamiliar, at least in the non-microbial world. The players are
cell-like entities still in early stages of their evolutions.  The
evolutionary dynamic (the \lq\lq rules") involve communal descent.  The
key element in this dynamic is {\it innovation-sharing\/}, an
evolutionary protocol whereby descent with variation from one \lq\lq
generation" to the next is not genealogically traceable, but is a
descent of a cellular community as a whole.  Even if an organismal
ancestry were to some extent traceable, it would have no significance,
because it is the community {\it as a unit\/}, not the individual
organismal lineages therein, that vary in descent.

The purpose of this paper is to show that evolvability, universality
and optimality can all be understood naturally and comprehensively, but
not within a framework of strictly vertical evolution. Specifically, we
will herein model the evolution of translation, the codon table, the
constraints therein, the universality of the code and the decoding
mechanism, not as a sum of parts, but as a whole.  This paper is
the first in a series that attempts to present such a comprehensive
model: subsequent articles will focus more on the decoding mechanism,
but here we focus primarily on innovation-sharing.  The central
conjecture in our model is that innovation sharing, which involves
horizontal transfer of genes and perhaps other complex elements among
the evolving entities---a dynamic far more rampant and pervasive than
our current perception of horizontal gene transfer---is required to
bring the evolving translation apparatus, its code and by implication
the cell itself to their current condition.

{\it Our point of view alleviates the need for any assumption of a
unique common ancestor.} We argue that the universality of the code is
a {\em generic consequence} of early communal evolution mediated by
horizontal gene transfer (HGT), and that HGT enhances optimality.  Our
arguments are backed up by computer simulation studies, which are
necessary in order to probe the complex interactions between the
variety of collective mechanisms that we shall present. We show that
there are virtuous cycles of cooperativity: 1) the more similar the
genetic codes are, the greater the intensity of horizontal gene
transfer is, and the stronger is the tendency for codes to become more
similar; 2) horizontal gene transfer helps the codes to optimize, and
optimization enforces universality and compatibility between
translational machineries. These cooperative dynamics arise because of
the dual role played by the genetic code: it is not only a protocol for
encoding amino acid sequences in the genome but also an
innovation-sharing protocol. Here we identify two synergistically
interacting mechanisms for the emergence of a universal
innovation-sharing protocol: dynamic competition between protocols
favoring the popular ones and effective attraction of codes due to
exchange of protein coding regions.  Other mechanisms are also
possible, and will be discussed elsewhere.

If Darwin had been a microbiologist, he surely would not have pictured
a \lq\lq struggle" for existence as \lq\lq red in tooth and claw". Our
view of competition in a communal world as a dynamical process is very
different from the widely-understood notion of Darwinian evolution.
\lq\lq Survival of the fittest" literally implies that there can only
be one winner from the forces of selection, whereas in a communal
world, the entire distributed community benefits and its structure
becomes modified by the forces of a selection that is an inherently
biocomplex phenomenon involving the dynamics between the community
elements and the interaction with the environment.  The most general
sense in which we mean competition in this article is the complex
dynamical rearrangement of the community structure.

Our framework fits naturally the recently proposed picture that early
evolution was dominated by horizontal gene transfer, as evidenced by
detailed phylogenetic\cite{WOES00}, biochemical\cite{SAUE05} and
structural\cite{ODON05} analyses of the aminoacyl-tRNA synthetases. The
broader implication of this scenario is that innovation sharing led to
the emergence of modern cell designs \cite{Woese2002} from a communal
state - not a unique, shared ancestor.  Such a communal state existed
prior to the point of emergence of vertical evolution, which has been
termed the Darwinian transition\cite{Woese2002}.  The defining property
of the communal state was that it was capable of tolerating and
utilizing ambiguity, as reflected in the pervasive role of horizontal
gene transfer.  A Darwinian transition corresponds to a state of
affairs when sufficient complexity has arisen that the state is
incapable of tolerating ambiguity, and so there is a distinct change in
the nature of the evolutionary dynamics---to vertical descent.  We
envision that such Darwinian transitions occurred in each of the three
major lineages. The present work does not address the Darwinian
transition itself, but explains how the communal state could have
arisen in the first place: in our scenario, it is the inevitable
by-product of the establishment of an innovation-sharing protocol---the
genetic code---leading to the explosive growth of complexity.  Thus, we
may speculate that the emergence of life should best be viewed in three
phases, distinguished by the nature of their evolutionary dynamics.  In
the first phase, treated in the present paper, life was very robust to
ambiguity, but there was no fully-unified innovation-sharing protocol.
The ambiguity in this stage led inexorably to a dynamic from which a
universal, and optimized innovation-sharing protocol emerged, through a
cooperative mechanism.  In the second phase, the community rapidly
developed complexity through the frictionless exchange of novelty
enabled by the genetic code---a dynamic we recognize to be patently
Lamarckian\cite{BURK77}.  With the increasing level of complexity there
arose necessarily a lower tolerance of ambiguity, leading finally to a
transition to a state wherein communal dynamics had to be suppressed
and refinement superseded innovation. This Darwinian transition led to
the third phase, which was dominated by vertical descent and
characterized by the slow, and tempered accumulation of complexity.

\section{Universality and HGT}

Previous arguments about universality rely on the existence of a
universal common ancestor with a frozen code. A detailed deconstruction
of such arguments is presented in the Supplementary Material, and
further supported by the computer simulations presented in Section
(\ref{sim}), but the unambiguous conclusion is that vertical descent on
its own is insufficient to explain the universality of the genetic
code. Here we present an alternative: the universality of the genetic
code is a generic consequence of the communal evolution of early life.
HGT of protein coding regions and HGT of translational components
ensures the emergence of clusters of similar codes and compatible
translational machineries. Different clusters compete for niches, and
due to the benefits of the communal evolution, the only stable solution
of the cluster dynamics is universality. Within clusters, concerted
optimization of codes is possible. These mechanisms are consistent with
two macroevolutionary scenarios: 1) the code stayed nearly universal at
all times, 2) The codes diverged at first but then gradually became
universal.

\subsection{Competition between innovation pools}
\label{popularity_contest}

One of the advantages of communal evolution is that universally good
traits and refinements can spread through HGT to organisms occupying
different niches, preserving their diversity. In a world increasingly
dominated by protein, most innovations would involve them, and
correspondingly HGT will be most effective between organisms having the
same genetic code. In this way, the organisms sort into communities
sharing related genetic codes. A single code community can span cells
adapted to different niches and with different organization.

The larger the community and diversity of organisms sharing the same
code, the larger is the pool of protein innovations accessible to
everyone. This leads to faster evolution among the larger communities
than the smaller ones, and therefore a greater potential to invade
niches occupied by organisms with different incompatible genetic codes.
With this dynamics larger communities will tend to become even larger
at the expense of smaller ones. The only stable solution is a
universal genetic code. Thus, it is not better genetic codes that give
an advantage but more common ones.

The elementary step in this process is the overtaking of an occupied
niche by the descendants of an organism with a different genetic code.
If two groups of organisms compete with each other, the one that has
access to more innovations (the one belonging to the larger community
of common/compatible genetic codes) will on average out compete the
other.  In contrast to the case with only vertical evolution, there is
an active feedback loop, driven by innovation sharing through HGT,
which not only singles out the genetic code from all other properties
of a cell, but also provides a mechanism that drives competition
between codes.

This mechanism---referred to below as \lq\lq competition between
innovation pools" or CIP---assumes that the protocols are fixed.
But how did the protocols themselves emerge and evolve?  How can a
protocol be upgraded without destroying it?

\subsection{Code attraction and optimization due to HGT of protein coding regions}

A population of organisms occupying a niche is subject to spontaneous
code mutations, and is bombarded by foreign genetic material from
organisms occupying different niches. Horizontally transferred genes
can be useful for the recipient even if the donor has a (somewhat)
different code.  For example, the codon usage (e.g. synonym codon usage
frequency) of a transferred gene is adapted to the donor code and
therefore different from that of the recipient. Correspondingly, there
will be indirect pressure for the recipient code to readjust itself to
make a better use of the new gene.

We expect that the code response would involve several characteristic
time scales.  On the long scale, the direction of change is to reduce
ambiguity, but on the short scale the code must be able to tolerate a
greater level of ambiguity while ingesting new genes. The means
available to the cell to detune the level of precision of translation
may be considered to be of two essential origins: those internal to the
cell, and those which are communal, reflecting the influence of the
environment and neighbouring cells.  Mechanisms internal to the cell
include change in tRNA expression levels and detuning of the ribosomal
machinery itself, as is known to occur through variations in Mg ion
concentration, antibiotics, and structural mutations.  Communal
mechanisms are likely to involve the import of tRNA from other
organisms.  The increase on the short time scale in translational
ambiguity is compensated for on the same time scale by the beneficial
effects of the new gene.  Eventually the codon and amino acid usage of
the newly transferred segment will equilibrate with the rest of the
genome and the indirect pressure of the donor code on the recipient
code will disappear, while leaving behind its accumulated effects.

In somewhat more detail, these arguments indicate that, after a HGT
event, the genetic code of the donor influences the genetic code of the
acceptor.  Given an alien gene, the host-alien gene system undergoes a
cyclical dynamical process leading to full utilization of the new gene.
In one part of the cycle, the host detunes its own code for purposes of
recognizing the alien code; an example of such a detuning process has
been documented in streptomycin-dependent mutants\cite{DAVI64, GARV73}
and ribosomal ambiguity mutants\cite{ROSS69, GORI70} in bacteria.  In
the other part of the cycle, the alien gene codons are mutated to
conform to the host code. This process results in the detection with
greater precision of the alien signal. A snapshot of this process would
reveal a genome as a mosaic of horizontally transferred fragments from
other genomes with different characteristic patterns of codon usage.
However, these are only the tip of the iceberg: beyond codon usage are
the subtle but important changes in amino-acyl tRNA synthetase
precision and the ambiguity level of the translational mechanism
itself.

The interaction between the genetic codes is attractive. Typically,
the closer the translation of a foreign coding region is to that in
the donor, the higher is the probability that it is functional.
Therefore, the selective pressure will be to change codons of the
recipient code in the direction of the donor code, even if only in a
probabilistic fashion. The dynamical outcome of this attraction must
be uniformization. This expectation is confirmed by the computer
simulations presented below.

HGT requires that the genetic codes of the host and the recipient are
sufficiently similar, but how similar is sufficient depends on the
nature of the proteins and the overall accuracy of decoding. There are
strong reasons to believe that the more primitive the code of the donor
is, the greater is the genetic code distance over which HGT is
possible. This is because the tolerance of the proteins to errors in
their primary structure is coadapted to the error rates of the
translational machinery. A cell with a non-optimal code cannot afford
very capricious and therefore highly fine-tuned proteins because of the
cost of discarding defective proteins. A protein that is robust to
translational errors {\it a fortiori\/} is also more tolerant to
translation with a different code. Conversely, the less optimized the
recipient code is, the more error-tolerant its proteins are, and
therefore the less harmful will be the effect on the established genes
of a code change in the direction of the donor code. This has the
important consequence that in the initial stages of the genetic code
evolution, when the diversification tendency of codes was strongest,
HGT was possible and must have been extensive despite the presence of
many different codes.

\subsection{HGT of translational components}

To this point, our discussion has managed to avoid the specifics of how
the genetic code is implemented in hardware, as it were.  However, we
cannot ignore the possibility that the translational components
themselves benefited from HGT, and we now turn to this briefly.  A
fuller account will be given elsewhere.

The genetic code is a representation of a family of modules, which are
universal across all organisms, and are specified by the mechanisms of
translation, such as tRNAs and charging enzymes (aminoacyl-tRNA
synthetases in a modern day setting). The task of improving translation
and the code is also universal, i.e. largely insensitive to the niches
organisms are occupying.  So is it possible that {\em HGT of
translational components} played an important role in the evolution of
the codes? Is there any significance in the functional separation
between the translational machinery (the ribosome) and the code
specificators (tRNAs and charging enzymes)? Imagine for simplicity a
situation in which organisms occupying diverse niches have the same
malleable genetic code and ensembles of tRNAs. The discovery of a tRNA
modification that changes the code and increases its optimality (and
therefore the efficiency of translation) in one organism will also be
beneficial for organisms in the other niches, due to the universal
benefit of optimality. Therefore, a spread of the discovery is
beneficial to all recipients, and can be assumed to occur through
various HGT mechanisms including via active elements such as viruses
and plasmids.

Therefore, if the spread through HGT is rapid compared to innovations,
a core of organisms having the same genetic code can maintain its
integrity while evolving towards optimality. Notice that this mechanism
does not rely on common ancestry and preserves the diversity of the
organisms.  Moreover, this mechanism is distinct from any survival of
any \lq\lq fittest" species.  In the absence of an attractive force
that restrains deviant codes, this core of organisms would become
depleted, if there was any circumstance that prevented a code update
from invading specific populations. If the depletion is slow enough,
the deviants will be at a communal disadvantage and disappear as
described in section \ref{popularity_contest}. The depletion mechanism
will compete against an expansion of the core due to the benefit of a
common protocol shared by a large population.

\subsection{Diversification of the translation mechanism}

The special role of the genetic code as an innovation-sharing protocol
leads to a possible observational consequence.  In a core community of
organisms that is in the process of code optimization, the
compatibility of code specificators is enforced. Once the optimization
of the genetic code is complete, there is no pressure to maintain
compatibility. Therefore the \lq\lq freezing" of the universal genetic
code could trigger the radiation of the underlying translational
machineries. So, even if translation emerged earlier than the other
basic cellular systems, but the optimization of the code took an
extended time, the translational componentry would have diversified
less. This is consistent with the observation that the translation
mechanism is more conserved evolutionary than the replication and
transcription ones.  Although we do not have a complete understanding
of the Darwinian transition \cite{Woese2002}, our argument suggests
that code universality and optimality were necessary but not sufficient
mechanisms for the transition to vertical evolution.

\subsection{Interactions between HGT mechanisms}

The different collective mechanisms enabled by HGT and outlined above
are also capable of synergistically interacting with each other.

We saw above that the evolutionary expansion of the most popular
cluster of codes provides the necessary support for the maintenance
of an otherwise weakly depleting universal core. The opposite is
also true. The CIP mechanism is ineffective if there are no clusters
of sufficient size on which it can operate. The establishment of
such clusters is greatly facilitated by the HGT of code
specificators and protein coding regions. Distribution of modules
enforces modularity that in turn enforces the distribution of
modules. Similarly, exchange of protein coding regions enforces
universality, thus making it easier to exchange genes. Therefore,
there are positive feedback loops that provide at least {\em local
stability} to the protocols, and turn them into effective degrees of
freedom at a longer time scale. The {\em global stability} and
universality is then guaranteed by the \lq\lq winner takes all"
nature of the CIP.

HGT of code specificators and protein coding regions interact not only
through the CIP mechanism but directly as well. If an organism obtains a gene
from another niche, its place in the ecosystem is such that it has
potential contact with the genetic material of the donor. Therefore,
the recipient has a better than random chance to obtain the right code
specificator from the donor as well, before the special codon usage of
a recently acquired trait drifts. The exchange of code specificators
provides a channel through which codes can become more similar in
response to the attraction of codes due to exchange of protein coding
regions.

In summary, it is the interaction between the different mechanisms
outlined above that makes the emergence and maintenance of universality
robust. At the same time, due to the complexity of the problem it is
useful to study the different components in isolation as well.

\subsection{HGT and the observed statistical properties of the genetic code}

So far, we have argued that HGT and the special role of the genetic
code as an innovation-sharing protocol alleviate the conceptual
difficulties in understanding the simultaneous universality and
evolvability of the genetic code. Does this improved understanding help
us explain some of the statistical features of the modern genetic code?
And how can we expose the signatures of the above mechanisms that are
buried in the functional and structural design of the translational
system and its phylogenetic variations?

To address this, one needs to complement the above generic mechanisms
with insight about the elementary evolutionary changes of the genetic
code.  Our goal in the remainder of this paper is to attempt to
identify robust or generic statistical properties of translation that
arise from our proposed evolutionary mechanisms, but which are
relatively insensitive to fine details. To begin, we model the code
attraction mechanism and ask: what is the effect of HGT on the
optimality of the genetic code?

We employ genetic code dynamics similar to that first introduced by
Sella and Ardell \cite{Sella_Ardell__mut_load,
Sella_Ardell__no_accident, Sella_Ardell__redundancy}. The main feature
is the coevolution between the genetic code and codon usage at
different functional sites. The code determines the codon usage at
mutation selection equilibrium. In turn, the codon usage determines the
fitness costs or benefits of the accessible code changes, thus guiding
the code's evolution. Code changes that are beneficial given the
typical codon usage of a population can invade it. To account for HGT
we couple the evolution of different codes by postulating that a
fraction of each genome consists of pieces coming from other genomes.

The virtue of Sella and Ardell's model is that it is a closed model of
the evolution of the genetic code, and shows that the evolvability
barrier is surmountable in a protein dominated world. Its shortcoming
is that it does not address the fact that translation is a dynamical
process, with competition between its various components. This means
that the Sella and Ardell model on its own is not adequate to identify
generic statistical signatures of the evolutionary mechanisms of
translation, because the statistical properties of the code and the
structure of the translational system are precisely the stable
resolutions of the design tradeoffs and evolutionary conflicts inherent
to translation.

The code attraction mechanism that we use at this point is also
insensitive to the implementation of translation, and so has the same
shortcoming. Thus, combining it with Sella and Ardell's model we will
still not be able to address all evolutionary aspects of the problem.
Nevertheless, such a model, while admittedly too simple for our
ultimate goals, can, encouragingly, still explain the universality and
optimality of the genetic code. The only key aspect of translation that
it is necessary to incorporate, even if by introducing it by hand, is
mistranslation.

At a next level in the hierarchy of models, one needs to incorporate
the tRNAs as agents of both the collective molecular effort of
translation and the communal evolution of the genetic codes. In
contrast to the above, in such models mistranslation would become a
dynamical variable that emerges from the competition itself.  This
lower level of description entails a richer suite of observational
outcomes, and provides a unique and essential role for the organism as
a resource manager and conflict regulator for the various dynamical
processes within it.

\section{Model of code attraction due to HGT}
\label{sim}

We first describe the modeling of the genetic code and
mistranslation errors, then define the proteome structure and the
reproductive success of genomes as a function of their codon
sequences. We show how to compute the probability distribution of
codons at different functional sites at mutation selection
equilibrium. Finally we present a simulation algorithm which
incorporates HGT.

\noindent {\em Genetic code:-\/} The genetic code is a probabilistic map
$Prob(c \rightarrow \alpha)$ between codons and amino acids. The
map is probabilistic because the charging of tRNAs with particular
amino acids and the decoding of codons through competition of
tRNAs are probabilistic molecular events. Quite generally,
\begin{equation}
Prob(c \rightarrow \alpha)  = \sum_t{T_{ct} C_{t\alpha}},
\end{equation}
where $T_{ct}$ is the probability that a codon $c$ is read by tRNA
species $t$, and $C_{t\alpha}$ is the probability that it is charged
with amino acid $\alpha$. Assuming one-to-one mapping between tRNA
species and codons, equal concentrations of different tRNAs, and
ignoring mischarging we set
\begin{equation}
 Prob(c \rightarrow \alpha)
= \sum_{c'}T_{cc'}\delta_{aa(c'), \alpha},
\end{equation}
where the sum is over the codons and $T_{cc'}=\nu/9$ if $c$ and $c'$
are nearest neighbors, $T_{cc}=1-\nu$ and $T_{cc'}=0$ otherwise,
with $\nu$ being the {\em mistranslation rate}. (9 is the number of
neighbors for codons consisting of three letters and an alphabet of
size 4.) $aa(c)$ is a map between the codons and amino acids, which
will be referred to as the {\em code}.

\noindent {\em Genome and proteome structure:-\/} A genome is a sequence of
codons that is translated to an amino acid sequence. Each genome
position $x$ belongs to a {\em site type} $s(x)$. A site type
$s$ is characterized by the {\em fitness score} $W_{\alpha\,s}$ of
the different amino acids that can be present at that site. The
matrix $W$, together with the frequencies $\left\{L_s\right\}$ of
the different site types in the genome, constitutes the {\it
structure of the proteome}. Assuming that amino acid substitutions
at different genome positions have independent effects on fitness,
we construct the {\it proteome fitness score}
\begin{equation}
A\Bigl(\bigl\{\alpha(x)\bigr\}\Bigr)= \prod_x W_{\alpha(x), s(x)},
\end{equation}
where the product is over all genome positions and $\alpha(x)$ is
the amino acid at position $x$.

\noindent {\em Codon usage:-\/} Since different positions belonging to the same
site type are phenotypically indistinguishable in the model, we can
describe the genome by the matrix $\{u_{sc}\}$ that specifies the
frequency of codon $c$ among sites of type $s$.

\noindent {\em Genome fitness:-\/} Accommodating the probabilistic nature of
translation, we set the fitness of a genome to be the average of the
proteome fitness score over many translations, i.e. $f = \langle A
\rangle$. Since translations of different codons are independent
\begin{equation}
f \Bigl(\{c(x)\}, code \Bigr)= \prod_x \langle W_{\alpha(x), s(x)}
\rangle \equiv \prod_x{F_{s(x), c(x)}} \:.
\end{equation}
Putting everything together and switching from codon sequence to
codon usage representation, we end up with
\begin{equation}
f\left(code, \{u\}\right) =
\prod_c\prod_s\left\{\sum_{c'}{T_{cc'}W_{aa(c'),
s}}\right\}^{L_s\,u_{sc}} \: . \label{fitness}
\end{equation}

\noindent {\em Equilibration of codon usage:-\/} Given the matrix of
mutational effects $\{F_{sc}(code)\}$, defined above, what is the codon
usage $\{u_{sc}\}$ in an asexual population of an infinite size and
large genomes at a mutation selection equilibrium? Mutational pressure
is characterized by the matrix $M_{cc'}$ specifying the probability
that codon $c$ will mutate into codon $c'$ in one generation. It is
assumed independent of the site type and genome position. Any
mutational biases could be incorporated in $M_{cc'}$. Here we focus on
equally probable single nucleotide changes. In this case, $M$ is
specified by a single parameter $\mu$ which is the probability for a
mutation at a given site in one generation. Following Sella and Ardell
\cite{Sella_Ardell__mut_load}, the codon usage at a site of type $s$ is
given by the eigenvector corresponding to the largest eigenvalue of the
matrix
\begin{equation}
Q_{cc'}^{(s)} \equiv M_{cc'} F_{sc'} \:. \label{Q_matrix}
\end{equation}
The matrix $Q$ reflects the application of selection followed by
mutation.

The parameters of the model described above are: $N_s$ - the number of site types,
$N_a$ - the number of amino acids, the $N_a \times N_s$ matrix $W$,
a vector $\{L_s\}$, specifying the relative frequencies of the
different site types, the mutation rate $\mu$ and the mistranslation
rate $\nu$.

\noindent {\em Model dynamics:-\/} Now we consider an ensemble of
populations with different codes and present the dynamics:
\begin{enumerate}
\item There are $N$ entities, each with its own genetic code
$aa(c)$ and codon usage $u_{sc}$.
\item At each step an entity, the acceptor, and $K$ random donor
entities are chosen at random. The acceptor codon usage is updated
according to the rule
\begin{equation}
(1 - \frac{H}{K} \sum_{k=1}^K{p_k} )\,u_{si} + \frac{H}{K}\,
\sum_{k=1}^K p_k u_{si}^{(k)}\longrightarrow u_{si} \ ,
\end{equation}
where $u_{si}^{(k)}$ is the codon usage of donor $k$, and $p_k$ is
some measure of the compatibility between the donor and acceptor
codes expressing the probability of acceptance. Here, we study the
case with {\em no barrier to HGT} of coding regions, i.e. $p_k = 1$.
$H$ is the fraction of the acceptor genome that is a mosaic due to
HGT.
\item We attempt to change the code of the acceptor. We examine in
random order the possible {\em elementary changes} of the $code$
until we find one that is acceptable or exhaust all the
possibilities. We accept a candidate change if it increases or at
least preserves the fitness , calculated using the mosaic codon
usage $\{u_{sc}\}$ and equation \ref{fitness}. An elementary code
change reassigns a single codon to a different amino acid.
\item We equilibrate the acceptor codon usage by finding the
eigenvectors corresponding to the largest eigenvalues of the matrices
$Q^{s}$.
\item
We repeat the cycle.
\end{enumerate}

The CIP mechanism, which clearly facilitates
universality (and given enough time generically leads to
universality), is factored out from the simulations in order to
concentrate on the code attraction mechanism. Each evolving entity
in the ensemble can be thought of as a different \lq\lq species"
(or ecotype). While within each species the evolution proceeds
through invasions of code variants with higher fitness, the
different species are stable and their number is fixed, thus
blocking the CIP mechanism.

\section{Results}

\subsection{Genetic code coevolution towards optimality and
universality}

We evolved ensembles of codes with and without HGT and measured the
time evolution of the average distance between codes and the
distribution of optimality scores. We compare the optimality scores
with the corresponding distribution for randomly generated codes.
The ensemble of randomly generated codes is constructed by assigning
random amino acids to the codons. Initially all the codes are
identical, and the initial code is generated by randomly assigning
amino acids to the codons.

The average code distance is obtained by considering all pairs of
entities with equal weight. The code distance between two entities
is the Hamming distance, which counts the numbers of codons that
code for different amino acids.

We define the optimality score of a code as the average amino acid
similarity distance between neighboring codons
\begin{equation}
\sum_c\sum_{c'} {N_{cc'} S_{aa(c),aa(c')}} \ ,
\label{optimality_score}
\end{equation}
where $N_{cc'}$ is 1 if codons $c$ and $c'$ differ by a single
letter and zero otherwise. $S$ is an the amino acid similarity
matrix defined as follows:
\begin{equation}
S_{\alpha\beta} = \sum_s{|W_{\alpha,s} - W_{\beta,s}|} \ .
\end{equation}

\begin{figure}
\includegraphics[width=\linewidth]{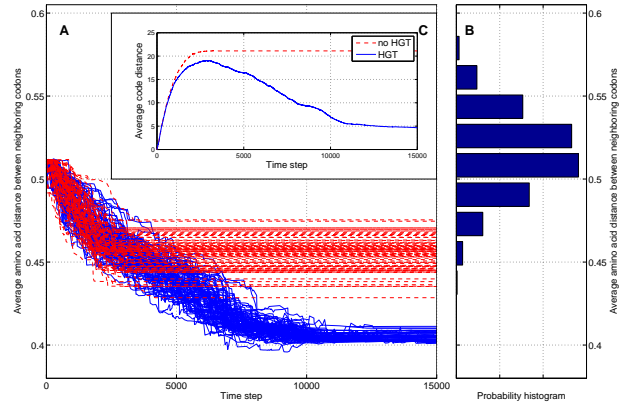}
\caption{Communal evolution towards optimality of $80$ codes with
(blue) and without (red) HGT of coding regions. There is no barrier to
HGT between different codes. The initial conditions are the same for
both runs. Parameters: $H=0.4$, $\phi=0.99$, $\mu=10^{-4}$, $\nu=0.01$.
(A) Time development of the average amino acid distance between
neighboring codons, a proxy for code optimality; (B) Probability
distribution histogram of code optimality for randomly-generated codes.
Horizontal axis is the frequency with which a given code optimality
occurs, vertical axis is the same as in (A); (C) Inset: time
development of the average distance between codes.} \label{fig1}
\end{figure}

Figure \ref{fig1} presents the simulation results for the following
parameters: $N=80$, $N_a=N_s=20$, $64$ codons, $\mu = 10^{-4}$, $\nu
= 0.01$, $W_{\alpha\beta}=\phi^{|A_{\alpha}-A_{\beta}|}$ with
$\{A_{\alpha}\}$ being uniformly distributed random numbers in the
interval $(0,1)$ and $\phi=0.99$. The HGT parameters are $H=0.4$ and
$K=1$.

The left panel - \ref{fig1}A, demonstrates that HGT of coding regions
not only brings universality but greatly enhances the joint ability of
the codes to optimize. Comparison with the distribution of optimality
scores for random codes, right panel - \ref{fig1}B, shows that, in the
presence of HGT, the achieved optimality is highly significant. Thus in
a qualitative way, we have provided a dynamical mechanism that would
give rise to the statistical properties of the genetic code identified
in \cite{Haig1991, One_in_a_million}.

The inset \ref{fig1}C shown that without HGT the codes diversify
form each other. However, when HGT is present the tendency to
diversify is eventually reversed and the codes get attracted to each
other, gradually achieving near universality. It should be stressed
that the probability for achieving universality, in the absence of
the CIP mechanism, depends on $H$, and is equal to
one above a threshold. While $H$ is a constant in this set of
simulations, the discussion in \cite{Woese2002} suggests that it is
in fact a dynamical variable that is initially large and gradually
decreases as better translation allows the evolution of a protein
network with more specific interactions.

We interpret these results as supporting two key concepts that underlie
the arguments in this paper.  First and foremost is the role of
communal evolution in leading to a universal genetic code.  Vertical or
Darwinian evolution does not lead to a reduction in the distance
between codes.  This is seen from the long time behavior of the red
curves in \ref{fig1}A.  Only the incorporation of HGT gives rise to
code convergence, as shown by the long time behaviour of the blue
curves in \ref{fig1}A: they get closer together with time.  In a sense,
a Darwinian (genealogical) evolution would get trapped, or perhaps,
frozen into metastable states.  Second is the role of communal
evolution in leading to an optimal code.  Vertical or Darwinian
evolution gets frozen into non-optimal states, whereas with HGT, the
code becomes optimized to a much greater extent.  This is seen by
comparing the final values of the Darwinian evolution (red) curves and
the communal evolution (blue) curves with the vertical axis of
\ref{fig1}B.  Communal evolution results in a genetic code that is much
further from the mean of random distributions than the results of
Darwinian evolution.

\section{Conclusions}

With this work, we have revisited the largely-overlooked problem of
genetic code universality and the conceptual difficulties associated
with it.  These difficulties can all be avoided if one takes, as we do,
the stance that evolution was essentially communal from the very
beginning.  We have argued that there are three distinct stages of
evolution, which we might classify as: (I) Weak communal evolution,
which gave way via development of an innovation-sharing protocol and
the emergence of a universal genetic code to (II) Strong communal
evolution, which developed exponential complexity of genes, finally
leading via the Darwinian transition to (III) Individual
evolution---vertical and so, Darwinian.

Most of our analysis explored the transition between regimes I and II,
through detailed consideration of the way in which a generalized form
of HGT operating on long evolutionary time scales brings universality
via dynamic competition between a wide variety of collective
innovation-sharing protocols.  In particular, we argued how such
protocols emerge through the important coevolutionary mechanism of code
attraction, and presented a specific model that is capable of
explaining the simultaneous universality and optimality of the genetic
code.

The genetic code is an expression of the translation process, and
therefore its state and significance reflect the various stages in the
evolutionary development of translation and the organization of the
cell.  Thus, a fuller account of the evolution of the genetic code
requires modeling physical components of the translational apparatus,
including the dynamics of tRNAs and the amino-acyl tRNA synthetases.
Only with this level of description is it possible to address issues,
such as the special role played by the polar requirement. This latter
point is, we believe, an essential clue to the early evolution of
translational components, when the genetic code presumably had a rather
different function.

Evolution of the genetic code, translation and cellular organization
itself follows a dynamic whose mode is, if anything, Lamarkian.

\section{Acknowledgments}

We would like to thank Patrick O'Donoghue, Gary Olsen, Yoshi Oono, Zan
Luthey-Schulten and Matt Gordon for many insightful discussions.  This
work was supported in part by the National Science Foundation through
grant number NSF-0526747.

\bibliographystyle{ieeetr}
\bibliography{refs}

\begin{thebibliography}{10}

\bibitem{WOES77}
C.~R. Woese and G.~E. Fox, ``The concept of cellular evolution,'' {\em J. Mol.
  Evol.}, vol.~10, pp.~1--6, 1977.

\bibitem{WOES82}
C.~R. Woese, ``Archaebacteria and cellular origins: an overview,'' {\em Zbl.
  Bakt. Hyg., I. Abt. Orig. C}, vol.~3, pp.~1--17, 1982.
\newblock Reprinted in Archaebacteria, O. Kandler (ed.), Fischer Verlag,
  Stuttgart, FRG, 1982.

\bibitem{Woese1965}
C.~Woese, ``On the evolution of the genetic code,'' {\em Proc. Natl. Acad.
  USA}, vol.~54, pp.~1546--–1552, 1965.

\bibitem{Woese1966}
C.~Woese, D.~Dugre, S.~Dugre, M.~Kondo, and W.~Saxinger, ``On the fundamental
  nature and evolution of the genetic code,'' {\em Cold Spring Harbour Symp.
  Quant. Biol.}, vol.~31, pp.~723--–736, 1966.

\bibitem{WOES-book}
C.~R. Woese, {\em The genetic code}.
\newblock New York: Harper and Row, 1967.

\bibitem{Sonneborn}
T.~Sonneborn, ``Degeneracy of the genetic code: Extent, nature and genetic
  implications,'' in {\em Evolving Genes and Proteins} (V.Bryson and H.~J.
  Vogel, eds.), (New York), pp.~277--297, Academic Press, 1965.

\bibitem{Alff-Steinberger1969}
C.~Alff-Steinberger, ``The genetic code and error transmission,'' {\em Proc.
  Natl. Acad. Sci. USA}, vol.~64, pp.~584--591, 1969.

\bibitem{WONG80}
J.~T.-F. Wong, ``Role of minimization of chemical differences between amino
  acids in the evolution of the genetic code,'' {\em Proc. Natl. Acad. USA},
  vol.~77, pp.~1083--1086, 1980.

\bibitem{Haig1991}
D.~Haig and L.~Hurst, ``A quantitative measure of error minimization in the
  genetic code,'' {\em J. Mol. Evol.}, vol.~33, pp.~412--–417, 1991.

\bibitem{One_in_a_million}
S.~Freeland and L.~Hurst, ``The genetic code is one in a million,'' {\em J.
  Mol. Evol.}, vol.~47, pp.~238--–248, 1998.

\bibitem{GOLD93}
N.~Goldman, ``Further results on error minimization in the genetic code,'' {\em
  J. Mol. Evol.}, vol.~37, pp.~662--664, 1993.

\bibitem{Freeland2000}
S.~Freeland, R.~Knight, L.~Landweber, and L.~Hurst, ``Early fixation of an
  optimal genetic code,'' {\em Mol. Biol. Evol.}, vol.~17, pp.~511--–518, 2000.

\bibitem{Knight-Nat_Rev_Genet}
R.~Knight, S.~Freeland, and L.~Landweber, ``Rewiring the keyboard: evolvability
  of the genetic code,'' {\em Nat. Rev. Genet.}, vol.~2, pp.~49--58, 2001.

\bibitem{Knight_PhD}
R.~Knight, {\em The origin and evolution of the genetic code: statistical and
  experimental investigations}.
\newblock PhD thesis, Princeton University, 2001.

\bibitem{WOES00}
C.~R. Woese, G.~Olsen, M.~Ibba, and D.~S\"oll, ``Aminoacyl-trna synthetases,
  the genetic code, and the evolutionary process,'' {\em Microbiology and
  Molecular Biology Reviews}, vol.~64, pp.~202--236, 2000.

\bibitem{SAUE05}
A.~Sauerwald, W.~Zhu, T.~Major, H.~Roy, S.~Palioura, D.~Jahn, W.~Whitman,
  J.~Yates, M.~Ibba, and D.~S\"oll, ``Rna-dependent cysteine biosynthesis in
  archaea,'' {\em Science}, vol.~307, pp.~1969--1972, 2005.

\bibitem{ODON05}
P.~O'Donoghue, A.~Sethi, C.~R. Woese, and Z.~A. Luthey-Schulten, ``The
  evolutionary history of cys-trna cys formation,'' {\em Proc. Natl. Acad. Sci.
  USA}, vol.~102, pp.~19003--19008, 2005.

\bibitem{Woese2002}
C.~Woese, ``On the evolution of cells,'' {\em Proc. Natl. Acad. Sci. USA},
  vol.~99, pp.~8742--8747, 2002.

\bibitem{BURK77}
J.~Richard W.~Burkhardt, {\em The spirit of system: Lamark and evolutionary
  biology}.
\newblock Harvard University Press, 1977.

\bibitem{DAVI64}
J.~Davies, W.~Gilbert, and L.~Gorini, ``Streptomycin, suppression and the
  code,'' {\em Proc. Natl. Acad. Sci. USA}, vol.~51, pp.~883--890, 1964.

\bibitem{GARV73}
R.~T. Garvin, R.~Rosset, and L.~Gorini, ``Ribosomal assembly influenced by
  growth in the presence of streptomyin,'' {\em Proc. Natl. Acad. Sci. USA},
  vol.~70, pp.~2762--2766, 1973.

\bibitem{ROSS69}
R.~Rosset and L.~Gorini, ``A ribosomal ambiguity mutation,'' {\em J. Mol.
  Biol.}, vol.~39, pp.~95--112, 1969.

\bibitem{GORI70}
L.~Gorini, ``Informational suppression,'' {\em Annu. Rev. Genet.}, vol.~4,
  pp.~107--134, 1970.

\bibitem{Sella_Ardell__mut_load}
G.~Sella and D.~Ardell, ``The impact of message mutation on the fitness of a
  genetic code,'' {\em J. Mol. Evol.}, vol.~54, pp.~638–--651, 2002.

\bibitem{Sella_Ardell__no_accident}
D.~Ardell and G.~Sella, ``No accident: genetic codes freeze in error-correcting
  patterns of the standard genetic code,'' {\em Phil. Trans. R. Soc. Lond. B},
  vol.~357, pp.~1625–--1642, 2002.

\bibitem{Sella_Ardell__redundancy}
D.~Ardell and G.~Sella, ``On the evolution of redundancy in genetic codes,''
  {\em J. Mol. Evol.}, vol.~53, pp.~269--–281, 2001.

\bibitem{Crick1968}
F.~Crick, ``The origin of the genetic code,'' {\em J. Mol. Biol.}, vol.~38,
  pp.~367--–379, 1968.

\bibitem{Wong1976}
J.~T. Wong, ``The evolution of a universal genetic code,'' {\em Proc. Natl.
  Acad. Sci. USA}, vol.~73, pp.~2336--2340, 1976.

\bibitem{Freeland_adaptation_for_adapting}
S.~Freeland, ``The darwinian genetic code: an adaptation for adapting?,'' {\em
  Genetic Programming and Evolvable Machines}, vol.~3, pp.~113--–127, 2002.

\end{thebibliography}

\section{Supplementary Material}

\subsection{Universality without HGT}

{\it Freezing before diversification:-} A possible sequence of events
leading to a frozen code Crick \cite{Crick1968} and Wong
\cite{Wong1976} is that translation emerged, the code optimized itself
and froze in a single niche, and only then did some other evolutionary
transition trigger the spread and diversification of its organisms. The
code is universal by virtue of competition and genetic drift between
organisms occupying this single niche. Or, perhaps, in a localized
ecosystem, there was more than one frozen code, but it was a single
lineage that diversified, conquering a world that was either unoccupied
or inhabited by inferior organisms lacking translation. The original
ecosystem was marginalized and the rest of the codes were
stochastically lost during the expansion. So, the universality of the
code is a result of a large scale \lq\lq founder" effect.

The problem with this {\em freezing before diversification} scenario is
that it does not explain what was stopping the expansion of organisms
endowed with some form of translation well before the genetic code
froze. The evolution of translation and the refinement of the genetic
code was most likely a multi-stage process that took an extended period
of time.

The alternative to {\em freezing before diversification} is {\em
freezing after diversification}. Even without understanding how codes
evolve, it is reasonable to assume that they eventually freeze.  Let's
look at the first moment of time when all organisms have frozen codes.
The codes diversified along with all other organism properties, so we
have many somewhat different codes and perhaps even some organisms that
do not have translation at all. How do we get from this situation to a
universal code? We now outline a variety of logically allowed
possibilities, explaining their merits and drawbacks.

{\it Stochastic universality:-} The universal genetic code is a
simple consequence of the fact that after an extended period of
time, all organisms will be descendants of a single organism and
will inherit its genetic code. This happens for stochastic reasons:
the descendants of an organism can invade a neighboring niche, and
by mere chance outcompete the organisms that were already there.
Since the entire phenotype space is connected, the repeated
stochastic takeover of neighboring niches will result eventually in
a universe inhabited by organisms with a common ancestor. The
problem here is that because of the stochastic nature of the process
it will be extremely slow. In addition the phenotype space is
constantly growing and it is not obvious that this slow stochastic
takeover process can ever saturate it.

{\it Evolutionary transition following translation:-} A universally
beneficial cellular property that emerged following the maturation
of translation caused a giant selective sweep overcoming the
preexisting adaptation of different organisms to different
environments \cite{Wong1976}. A candidate for such an event is the
discovery of DNA. This is an attractive possibility. One of the
problems is that by the time of maturation of translation we might
already have different cell designs with different ecological roles
and it is not immediately clear how even such a great innovation can
overcome this. In contrast, if the code was already universal due to
the communal evolution, proteins involved in DNA processing could
have been easily distributed.

{\it Selection on optimality:-} Once proteins were established as a
major determinant of the phenotype, the quality of the genetic code
became the single most important contributor to fitness.  The more
optimized genetic codes out-competed less optimized ones. Related is
the suggestion that optimality is linked to evolvability
\cite{Freeland_adaptation_for_adapting}, and therefore, organisms with
more optimal codes are evolutionary more successful. The problem here
is that it is not {\it a priori\/} clear that we can ignore the
diversity and competition along all the other phenotypic dimensions.
Moreover, the more optimized the codes are, the less will the
differences between them matter. So we would expect after all a
diversification of the genetic codes.

We conclude that, on rather general grounds, an explanation of code
universality based on vertical evolution is likely to be
problematic.

\end{document}